\documentclass{PoS}
\usepackage{epsfig}
\usepackage{graphics}

\title{Study of the $a_0(980)$ on the lattice}

\ShortTitle{Study of the $a_0(980)$ on the lattice}

\author{European Twisted Mass Collaboration (ETMC):}

\author{\speaker{Marc Wagner} \\
        Goethe-Universit\"at Frankfurt am Main, Institut f\"ur Theoretische Physik, \\ $\quad$ Max-von-Laue-Stra{\ss}e 1, D-60438 Frankfurt am Main, Germany}

\author{Abdou Abdel-Rehim, Constantia Alexandrou, Mario Gravina, Giannis Koutsou \\
        Department of Physics, University of Cyprus, P.O.\ Box 20537, 1678 Nicosia, Cyprus \\
        Computation-based Science and Technology Research Center, Cyprus Institute, 20 Kavafi Street, \\ $\quad$ Nicosia 2121, Cyprus}

\author{Mattia Dalla Brida \\
        School of Mathematics, Trinity College Dublin, Dublin 2, Ireland}

\author{Luigi Scorzato \\
        INFN -- gruppo collegato di Trento, via sommarive 14, 38123 -- Trento, Italy}

\author{Carsten Urbach \\
        Helmholtz-Institut f{\"u}r Strahlen- und Kernphysik (Theorie) and Bethe Center for Theoretical \\ $\quad$ Physics, Universit{\"a}t Bonn, D-53115 Bonn, Germany}

\abstract{
We present lattice results for the $a_0(980)$ state using a variational approach with a variety of creation operators: quark-antiquark, mesonic molecule, diquark-antidiquark as well as two-meson type. Focus is put on recent technical advances, in particular the computation of singly disconnected diagrams.
}

\FullConference{31st International Symposium on Lattice Field Theory, LATTICE 2013\\
		July 29--August 3, 2013\\
		Mainz, Germany}

\begin{document}


\section{Introduction}

The nonet of light scalar mesons formed by $\sigma \equiv f_0(500)$, $\kappa \equiv K_0^\ast(800)$, $a_0(980)$ and $f_0(980)$ is poorly understood \cite{Jaffe:2004ph}. Compared to expectation all nine states are rather light and their ordering is inverted. For example in a standard quark antiquark picture the $a_0(980)$ states, which have $I = 1$, must necessarily be composed of two light quarks, e.g.\ $a_0(980) \equiv \bar{d} u$, while the $\kappa$ states with $I = 1/2$ are made from a strange and a light quark, e.g.\ $\kappa \equiv \bar{s} u$. Consequently, $\kappa$ should be heavier than $a_0(980)$, since $m_s > m_{u,d}$. In experiments, however, the opposite is observed, i.e.\ $m(\kappa) = 682 \pm 29 \, \textrm{MeV}$, while $m(a_0(980)) = 980 \pm 20 \, \textrm{MeV}$ \cite{PDG}. On the other hand in a four-quark or tetraquark picture the quark content could be $a_0(980) \equiv \bar{d} u \bar{s} s$ and $\kappa \equiv \bar{s} u (\bar{u} u + \bar{d} d)$ naturally explaining the observed ordering. Moreover, certain decay channels, e.g.\ $a_0(980) \rightarrow K + \bar{K}$, indicate that besides the two light quarks also an $s \bar{s}$ pair is present and, therefore, also support a tetraquark interpretation.

Here we report about the status of an ongoing long-term project with the aim to study scalar mesons and other possible tetraquark candidates, in particular the $a_0(980)$, using lattice QCD. In section~\ref{SEC001} we summarize our recently published results obtained with Wilson twisted mass quarks, where ``singly disconnected diagrams'' (connected diagrams with closed fermion loops) have been neglected \cite{Daldrop:2012sr,Alexandrou:2012rm,Wagner:2012ay,Wagner:2013nta}. In section~\ref{SEC455} we discuss latest technical advances obtained with clover improved Wilson quarks, in particular the inclusion of singly disconnected diagrams.


\section{\label{SEC001}Wilson twisted mass quarks, singly disconnected diagrams neglected}


\subsection{\label{SEC388}Lattice setup}

We use gauge link configurations with $N_f = 2+1+1$ dynamical quark flavors generated by the ETM Collaboration \cite{Baron:2010bv,Baron:2010th}. We consider several ensembles with lattice spacing $a \approx 0.086 \, \textrm{fm}$. The ensembles differ in the volume $(L/a)^3 \times (T/a) = 20^3 \times 48 , \ldots , 32^3 \times 64$ and the unphysically heavy light quark mass corresponding to $m_\pi \approx 280 \, \textrm{MeV} \ldots 460 \, \textrm{MeV}$.

For the computations presented in this section we have ignored singly disconnected diagrams, which are technically more challenging than their connected counterparts (cf.\ section~\ref{SEC500}). An important physical consequence is that the quark number and the antiquark number are separately conserved for each flavor. Therefore, there is no mixing between $\bar{u} u$, $\bar{d} d$ and $\bar{s} s$ resulting in an $\eta_s$ meson with flavor structure $\bar{s} s$ instead of $\eta$ and $\eta'$. Moreover, there is no mixing between two-quark and four-quark states, which is why we consider in this section only four-quark creation operators and ignore ordinary quark-antiquark creation operators.


\subsection{Four-quark creation operators}

$a_0(980)$ has quantum numbers $I(J^P) = 1(0^+)$. As usual in lattice QCD we extract the low lying spectrum in that sector by studying the asymptotic exponential behavior of correlation functions $C_{j k}(t) = \langle (\mathcal{O}_j(t))^\dagger \mathcal{O}_k(0) \rangle$. $\mathcal{O}_j$ and $\mathcal{O}_k$ denote suitable creation operators, i.e.\ operators generating the $a_0(980)$ quantum numbers, when applied to the vacuum state.

Assuming that the experimentally measured $a_0(980)$ with mass $980 \pm 20 \, \textrm{MeV}$ is a rather strongly bound four quark state, suitable creation operators to excite such a state are
\begin{eqnarray}
\label{EQN001} & & \hspace{-0.7cm} \mathcal{O}_{a_0(980)}^{K \bar{K} \textrm{\scriptsize{} molecule}} \ \ = \ \ \sum_\mathbf{x} \Big(\bar{s}(\mathbf{x}) \gamma_5 u(\mathbf{x})\Big) \Big(\bar{d}(\mathbf{x}) \gamma_5 s(\mathbf{x})\Big) \\
\label{EQN002} & & \hspace{-0.7cm} \mathcal{O}_{a_0(980)}^{\textrm{\scriptsize diquark}} \ \ = \ \ \sum_\mathbf{x} \Big(\epsilon^{a b c} \bar{s}^b(\mathbf{x}) C \gamma_5 \bar{d}^{c,T}(\mathbf{x})\Big) \Big(\epsilon^{a d e} u^{d,T}(\mathbf{x}) C \gamma_5 s^e(\mathbf{x})\Big) .
\end{eqnarray}
The first operator has the spin/color structure of a $K \bar{K}$ molecule. The second resembles a bound diquark antidiquark pair, where spin coupling via $C \gamma_5$ corresponds to the lightest diquarks/antidi\-quarks (cf.\ e.g.\ \cite{Jaffe:2004ph,Alexandrou:2006cq,Wagner:2011fs}). Further low lying states with $a_0(980)$ quantum numbers to be considered are the two-particle states $K + \bar{K}$ and $\eta_s + \pi$. Suitable creation operators to resolve these states are
\begin{eqnarray}
\label{EQN003} & & \hspace{-0.7cm} \mathcal{O}_{a_0(980)}^{K + \bar{K} \textrm{\scriptsize{} two-particle}} \ \ = \ \ \bigg(\sum_\mathbf{x} \bar{s}(\mathbf{x}) \gamma_5 u(\mathbf{x})\bigg) \bigg(\sum_\mathbf{y} \bar{d}(\mathbf{y}) \gamma_5 s(\mathbf{y})\bigg) \\
\label{EQN004} & & \hspace{-0.7cm} \mathcal{O}_{a_0(980)}^{\eta_s + \pi \textrm{\scriptsize{} two-particle}} \ \ = \ \ \bigg(\sum_\mathbf{x} \bar{s}(\mathbf{x}) \gamma_5 s(\mathbf{x})\bigg) \bigg(\sum_\mathbf{y} \bar{d}(\mathbf{y}) \gamma_5 u(\mathbf{y})\bigg) .
\end{eqnarray}


\subsection{\label{SEC377}Numerical results}

We first discuss numerical results for the ensemble with the smallest volume, $(L/a)^3 \times (T/a) = 20^3 \times 48$, which corresponds to a spatial extension of $L \approx 1.72 \, \textrm{fm}$. This ensemble is particularly suited to distinguish two-particle states with relative momentum from states with two particles at rest and from possibly existing $a_0(980)$ tetraquark states.

\begin{figure}[htb]
\begin{center}
\includegraphics[angle=-90,width=7.0cm]{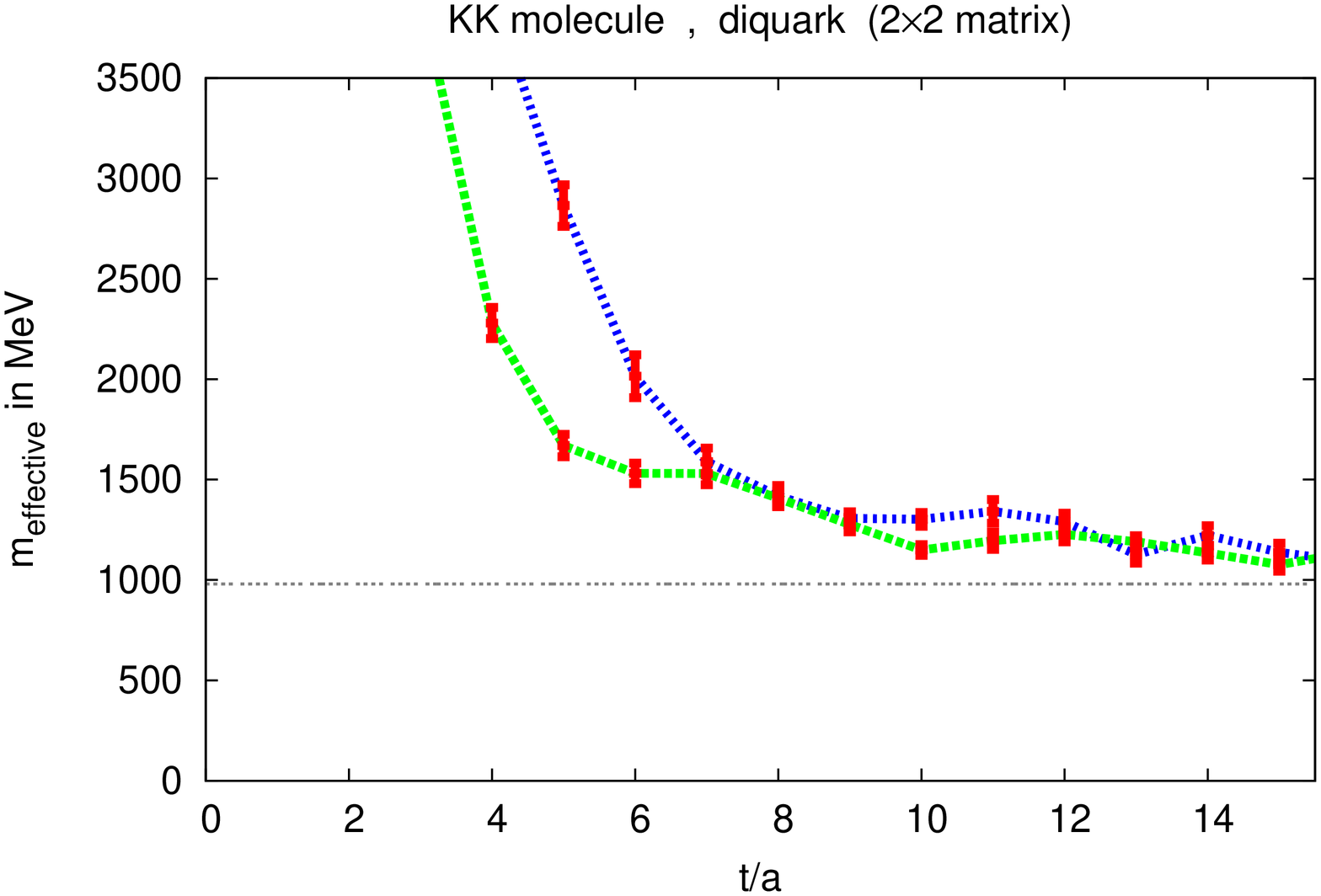}
\includegraphics[angle=-90,width=7.0cm]{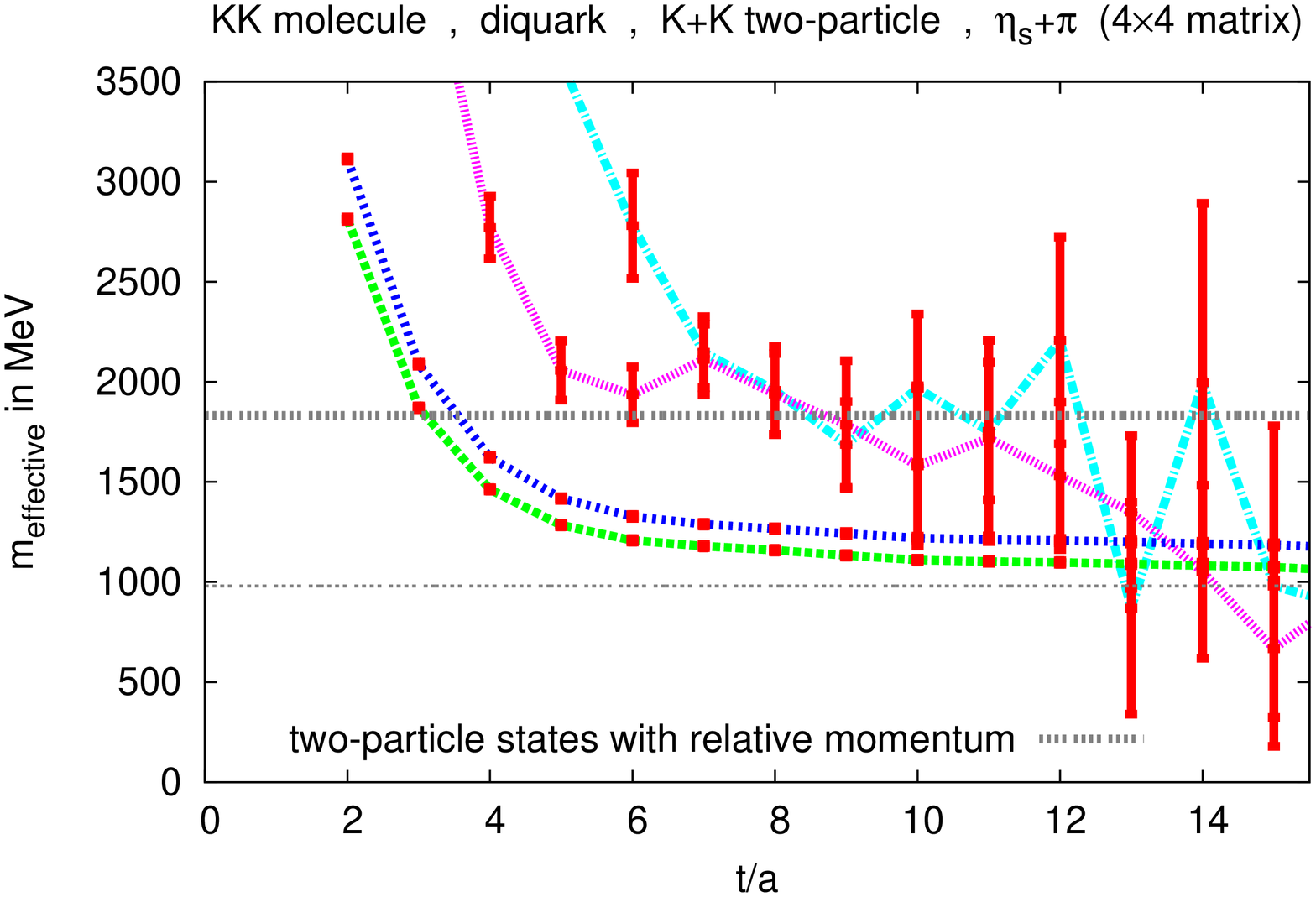}
\caption{\label{F001}Effective masses as functions of the temporal separation.
\textbf{(left)}~$2 \times 2$ correlation matrix (operators: $K \bar{K}$ molecule, diquark).
\textbf{(right)}~$4 \times 4$ correlation matrix (operators: $K \bar{K}$ molecule, diquark, $K + \bar{K}$ two-particle, $\eta_s + \pi$ two-particle).
}
\end{center}
\end{figure}

Figure~\ref{F001}(left) shows effective mass plots for a $2 \times 2$ correlation matrix with a $K \bar{K}$ molecule operator (\ref{EQN001}) and a diquark-antidiquark operator (\ref{EQN002}). The corresponding two plateaus are around $1100 \, \textrm{MeV}$ and, therefore, consistent both with the expectation for possibly existing $a_0(980)$ tetra\-quark states and with two-particle $K + \bar{K}$ and $\eta_s + \pi$ states, where both particles are at rest ($2 \times m(K) \approx 1198 \, \textrm{MeV}$; $m(\eta_s) + m(\pi) \approx 1115 \, \textrm{MeV}$ in our lattice setup).

Increasing this correlation matrix to $4 \times 4$ by adding $K + \bar{K}$ two-particle and $\eta_s + \pi$ two-particle operators (eqs.\ (\ref{EQN003}) and (\ref{EQN004})) yields the effective mass results shown in Figure~\ref{F001}(right). Two additional states are observed, whose plateaus are around $1500 \, \textrm{MeV} \ldots 2000 \, \textrm{MeV}$. From this $4 \times 4$ analysis we conclude the following:
\begin{itemize}
\item We do not observe a third low-lying state around $1100 \, \textrm{MeV}$, even though we provide operators, which are of tetraquark type as well as of two-particle type. This suggests that the two low-lying states are the expected two-particle $K + \bar{K}$ and $\eta_s + \pi$ states, while an additional stable $a_0(980)$ tetraquark state does not exist.

\item The effective masses of the two low-lying states are of much better quality in the right plot of Figure~\ref{F001} than in the left plot. We attribute this to the $K + \bar{K}$ two-particle and $\eta_s + \pi$ two-particle operators, which presumably create larger overlap to those states than the tetraquark operators. This confirms the interpretation of the two observed low-lying states as two-particle states.

\item We investigated these overlaps in a more quantitative way by studying the squared eigenvector components of the two low-lying states (cf.\ \cite{Alexandrou:2012rm} for details). The lowest state is predominantly of $\eta_s + \pi$ type, whereas the second lowest state is of $K + \bar{K}$ type. The two tetraquark operators are essentially irrelevant for resolving those states. This strongly supports the above interpretation of the two observed low lying states as two-particle states.

\item The energy of two-particle excitations with one relative quantum of momentum can be estimated: $m(K + \bar{K},p = p_\textrm{\scriptsize min}) \approx 1873 \, \textrm{MeV}$ and $m(\eta_s + \pi,p = p_\textrm{\scriptsize min}) \approx 1853 \, \textrm{MeV}$ \cite{Alexandrou:2012rm}. These values are consistent with the effective mass plateaus of the second and third excitation in Figure~\ref{F001}(right). Consequently, we also interpret them as two-particle states.
\end{itemize}

We obtained qualitatively identical results, when varying the light quark mass and the spacetime volume, using the ensembles discussed in section~\ref{SEC388}. Corresponding plots are shown in \cite{Alexandrou:2012rm}.


\section{\label{SEC455}Clover improved Wilson quarks, singly disconnected diagrams included}


\subsection{Lattice setup}

Recently we started to perform similar computations using gauge link configurations generated by the PACS-CS Collaboration \cite{Aoki:2008sm} with 2+1 flavors of clover improved Wilson sea quarks and the Iwasaki gauge action. A significant advantage compared to Wilson twisted mass quarks is that parity and isospin are exact symmetries. For example there is no pion and kaon mass splitting and $P = +$ and $P = -$ states are separated by quantum numbers (these problems in the context of Wilson twisted mass quarks and the $a_0(980)$ are discussed in detail in \cite{Alexandrou:2012rm}). The currently used ensemble has a lattice spacing $a \approx 0.09 \, \textrm{fm}$, a volume $(L/a)^3 \times T/a = 32^3 \times 64$ and light $u/d$ quarks corresponding to $m_\pi \approx 300 \, \textrm{MeV}$. 


\subsection{\label{SEC500}Singly disconnected contributions}

A major improvement compared to our previous results presented in section~\ref{SEC377} is that this time singly disconnected contributions to the correlation functions are included. These singly disconnected contributions correspond to diagrams, where the strange quark propagators start and end at the same timeslice. Ignoring such singly disconnected diagrams introduces a systematic error, which is hard to quantify. With these new computations, where disconnected contributions are taken into account, this systematic error has been eliminated. Another important consequence of singly disconnected diagrams is that quark-antiquark and four-quark trial states can have non-vanishing overlap. This allows to study a non-trivial larger correlation matrix containing both a quark-antiquark operator
\begin{eqnarray}
\label{EQN645} \mathcal{O}_{a_0(980)}^{q \bar{q}} \ \ = \ \ \sum_\mathbf{x} \Big(\bar{d}(\mathbf{x}) u(\mathbf{x})\Big)
\end{eqnarray}
and the four-quark operators (\ref{EQN001}) to (\ref{EQN004}). Such a correlation matrix will allow to make stronger statements about the structure of states from the $a_0(980)$ sector.

\begin{figure}[htb]
\begin{center}
\includegraphics[width=12.0cm]{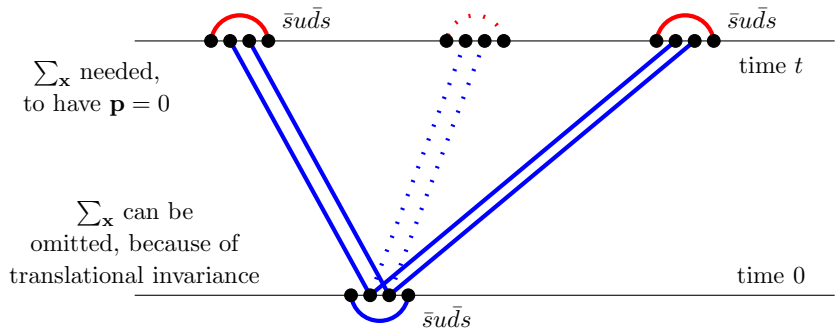}
\caption{\label{FIG002}The singly disconnected diagram of a $\bar{s} s \bar{d} u$ correlator.}
\end{center}
\end{figure}

The technique for computing singly disconnected diagrams presented in the following is quite general. Since here we are interested in the $a_0(980)$, we discuss it in the context of four-quark creation operators with flavor structure $\bar{s} s \bar{d} u$ (cf.\ Figure~\ref{FIG002}).

While for connected four-quark diagrams (i.e.\ all four quark propagators connect the timeslices at time $0$ and time $t$) standard point-to-all propagators can be used, applying exclusively such propagators is not possible in practice for singly disconnected diagrams. The reason is that one has to include a sum over space, $\sum_\mathbf{x}$, at least on one of the two timeslices (wlog.\ at time $t$ in Figure~\ref{FIG002}), to project to zero momentum. This in turn requires an all-to-all $s$ quark propagator on that timeslice, due to $\sum_\mathbf{x} \bar{s}(\mathbf{x},t) s(\mathbf{x},t) \ldots$ (the solid red lines in Figure~\ref{FIG002}).

Since all-to-all propagators are prohibitively expensive to compute, they are typically estimated stochastically. While using a single stochastic propagator for a specific diagram typically results in a favorable or at least acceptable signal-to-noise ratio, using a larger number of such propagators drastically increases statistical errors. Therefore, we decided for the following strategy: three quark propagators (the $s$-loop at timeslice $0$ and the $u$ and $d$ propagators connecting the timeslices $0$ and $t$) are realized by exact point-to-all propagators, while the remaining propagator (the $s$-loop at timeslice $t$) is estimated stochastically using random $Z_2 \times Z_2$ timeslice sources.

Numerical tests of this strategy have shown that the statistical errors of the singly disconnected diagrams are of the same order of magnitude as the statistical errors of the corresponding connected diagrams, when investing a comparable amount of HPC resources.

\subsection{First numerical results}

At the moment we have only performed computations of a $2 \times 2$ correlation matrix containing the $q \bar{q}$ operator (eq.\ (\ref{EQN645})) and the four-quark $K \bar{K}$ molecule operator (eq.\ (\ref{EQN001})). Both connected and singly disconnected diagrams have been computed using 111 gauge link configurations with a single sample on each configuration.

\begin{figure}[htb]
\begin{center}
\includegraphics[width=7.0cm]{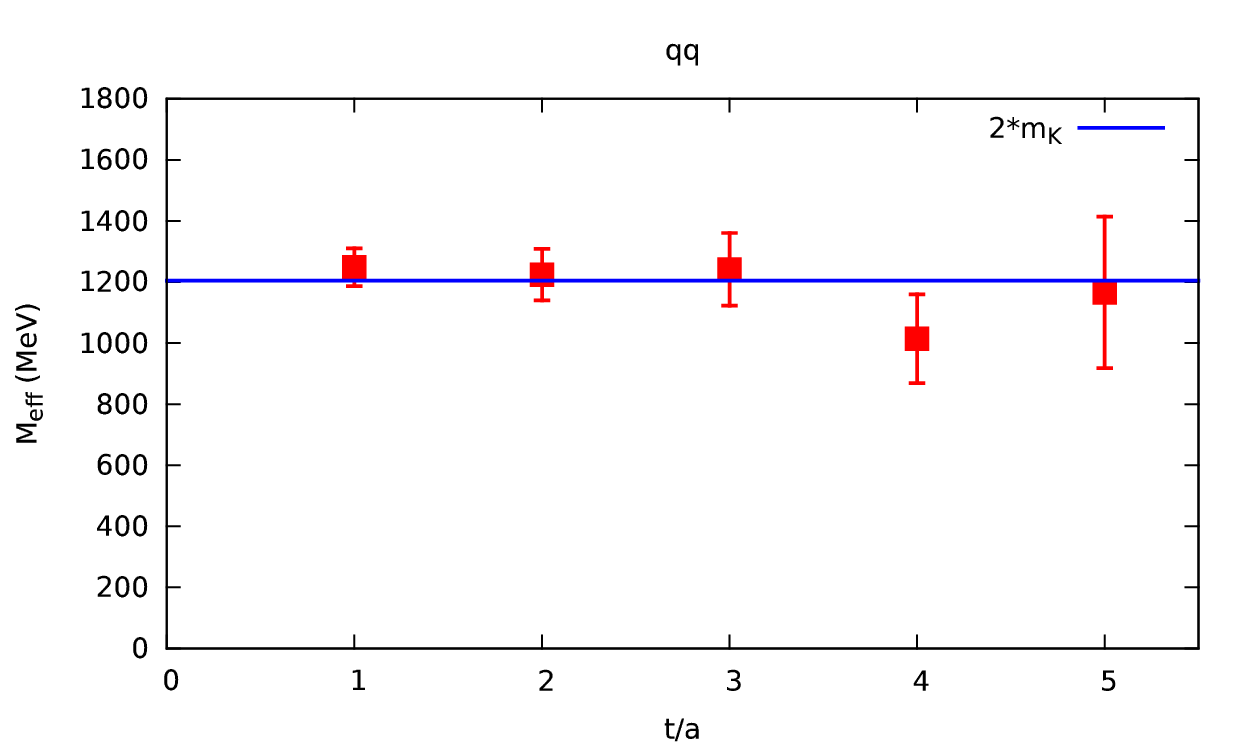}
\includegraphics[width=7.0cm]{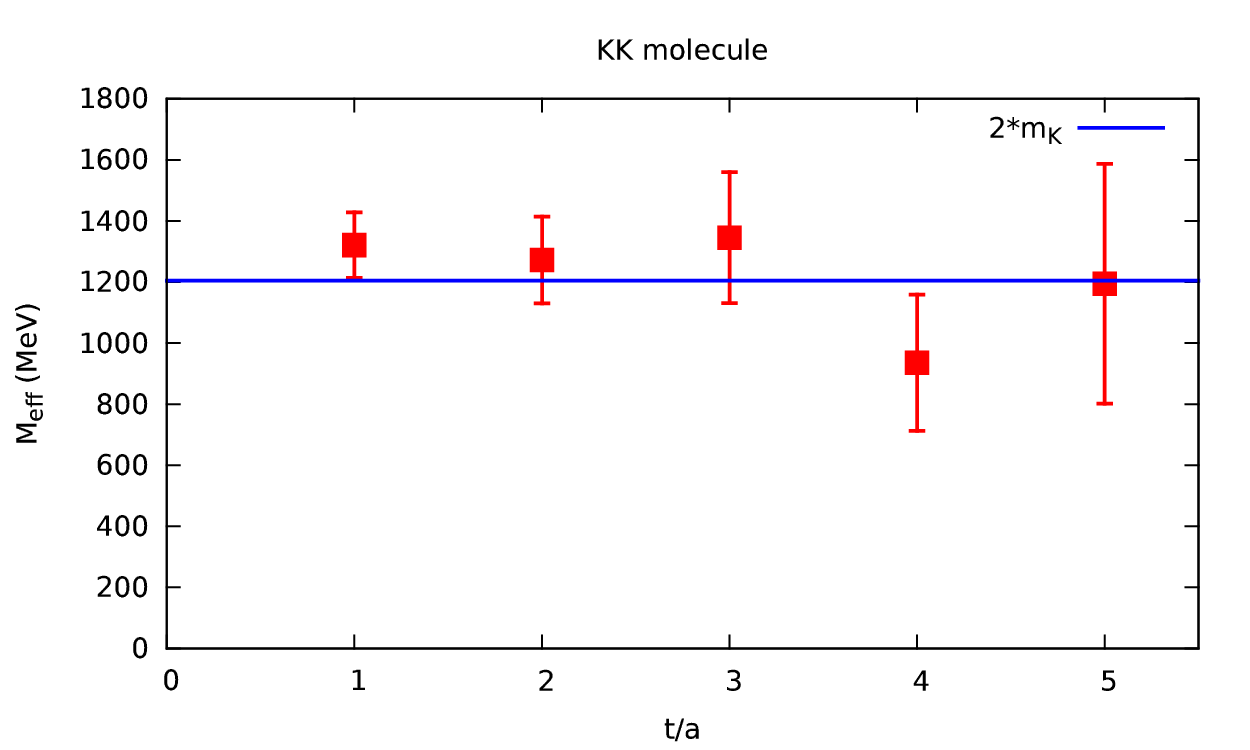}
\includegraphics[width=7.0cm]{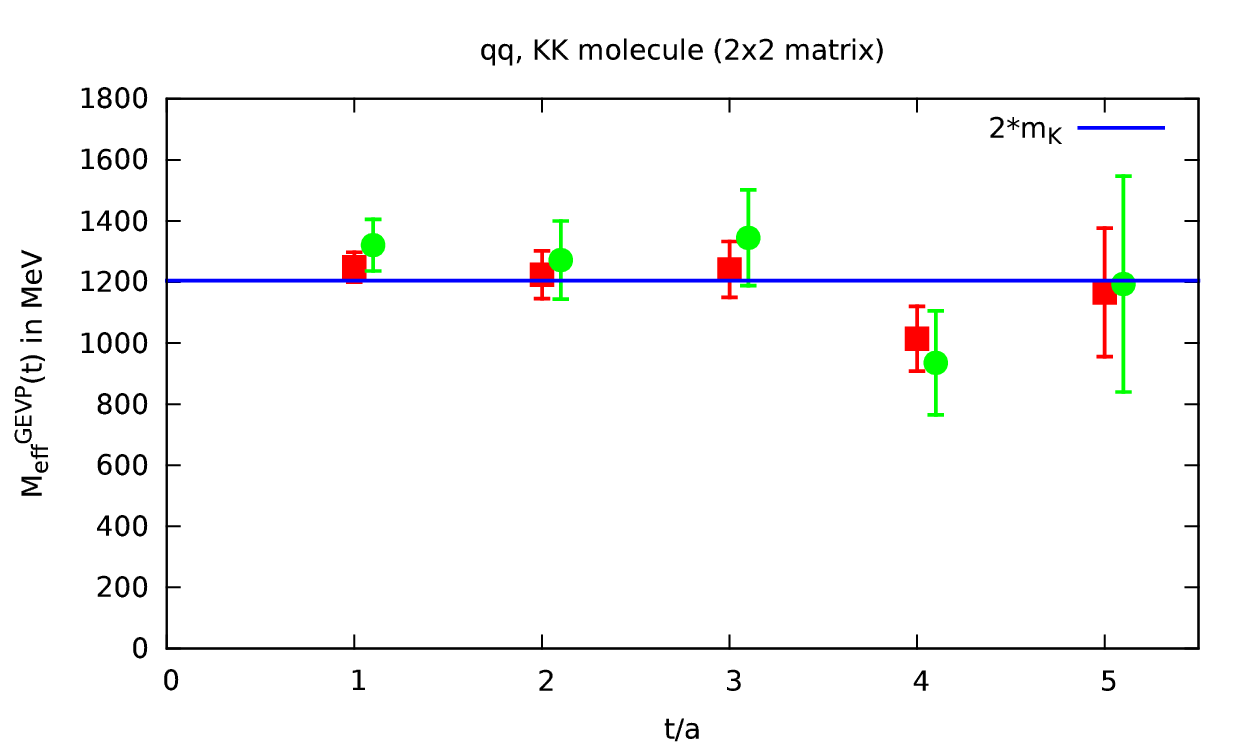}
\caption{\label{F004}Effective masses as functions of the temporal separation.
\textbf{(left top)}~$q \bar{q}$ operator.
\textbf{(right top)}~$K \bar{K}$ molecule operator.
\textbf{(bottom)}~$2 \times 2$ correlation matrix (operators: $q \bar{q}$, $K \bar{K}$ molecule).
}
\end{center}
\end{figure}

Effective masses corresponding to the two diagonal elements, i.e.\ individually to the $q \bar{q}$ operator and the $K \bar{K}$ molecule operator, are plotted in Figure~\ref{F004}, upper line. The plot in the lower line shows the two effective masses obtained by solving a generalized eigenvalue problem for the above mentioned $2 \times 2$ correlation matrix. All effective masses are around $1000 \, \textrm{MeV}$ and, therefore, consistent with $2 \times m(K)$, with $m(\eta) + m(\pi)$ and also with $m(a_0(980))$.

For statements regarding the structure of the observed states we need to include also the four-quark operators (\ref{EQN002}) to (\ref{EQN004}). This will allow us to address similar questions as in section~\ref{SEC377}, i.e.\ to answer, whether there is in addition to the two-particle $K + \bar{K}$ and $\eta + \pi$ states also a third (bound) state near the mass of $a_0(980)$.


\section{Future plans}

Our next step will be the computation of a $5 \times 5$ correlation matrix containing the creation operators (\ref{EQN001}) to (\ref{EQN004}) and (\ref{EQN645}), with singly disconnected diagrams included. This will allow us, to perform a similar analysis as in section~\ref{SEC377}, this time, however, without neglecting singly disconnected contributions.

Since the $a_0(980)$ is most likely a weakly bound unstable state, we also intend to study the volume dependence of the two-particle spectrum using ``L\"uscher's method'' (cf.\ e.g.\ \cite{Luscher:1986pf,Luscher:1990ux,Luscher:1991cf}) or improved techniques taking e.g.\ coupled channel scattering into account \cite{Bernard:2010fp,Doring:2012eu,Doring2013}. Such computations are very challenging using lattice QCD, but first results have recently been published, e.g.\ for the $\kappa$ and positive parity $D$ mesons \cite{Lang:2012sv,Mohler:2012na}. We plan to perform similar computations in the near future.


\begin{acknowledgments}

M.W.\ acknowledges support by the Emmy Noether Programme of the DFG (German Research Foundation), grant WA 3000/1-1. M.G.\ acknowledges support by the Marie-Curie European training network ITN STRONGnet grant PITN-GA-2009-238353. M.D.B.\ is funded by the Irish Research Council and is grateful for the hospitality at the University of Cyprus and Cyprus Institute, where part of this work was carried out. L.S.\ acknowledges support by STRONGnet. This work was supported in part by the Helmholtz International Center for FAIR within the framework of the LOEWE program launched by the State of Hesse and by the DFG and the NSFC through funds provided to the Sino-German CRC 110 ``Symmetries and the Emergence of Structure in QCD''.

\end{acknowledgments}



\end{document}